\documentclass[conference,a4paper]{IEEEtran}
\IEEEoverridecommandlockouts
\usepackage{cite}
\usepackage{amsmath,amssymb,amsfonts}
\usepackage{graphicx}
\usepackage{textcomp}
\usepackage{xcolor}
\usepackage{comment}
\usepackage{algorithm}
\usepackage{algpseudocode}
\usepackage{float}
\usepackage{bbm}

\usepackage{tikz,pgf}
\usetikzlibrary{patterns}
\usetikzlibrary{calc}

\usepackage{pgfplots}

\newtheorem{definition}{Definition}

\def\BibTeX{{\rm B\kern-.05em{\sc i\kern-.025em b}\kern-.08em
    T\kern-.1667em\lower.7ex\hbox{E}\kern-.125emX}}

\begin{document}

\title{``Fog" Optimization via Virtual Cells in \\ Cellular Network Resource Allocation
\thanks{This research was supported by AFOSR Grant FA9550-12-1-0215 and ONR Grant  
	N000141512527}
}

\author{\IEEEauthorblockN{ Michal Yemini}
\IEEEauthorblockA{Wireless Systems Laboratory\\
Stanford University\\
California, USA \\
michalye@stanford.edu}
\and
\IEEEauthorblockN{Andrea J. Goldsmith}
\IEEEauthorblockA{Wireless Systems Laboratory\\
Stanford University\\
California, USA \\
andrea@wsl.stanford.edu}

}

\maketitle

\begin{abstract}
This work proposes a new resource allocation optimization framework for cellular networks using ``fog" or neighborhood-based optimization rather than fully centralized or fully decentralized methods. In neighborhood-based optimization resources are allocated within virtual cells encompassing several base-stations and the users within their coverage area. As the number of base-stations within a virtual cell increases, the framework reverts to centralized optimization, and as this number decreases it reverts to decentralized optimization. We address two  tasks that must be carried out in the fog optimization framework: forming the virtual cells and allocating the communication resources in each virtual cell effectively. We propose hierarchical clustering for the formation of the virtual cells given a particular number of such cells. Once the virtual cells are formed, we consider several optimization methods to solve the NP-hard joint channel access and power allocation problem within each virtual cell in order to maximize the sum rate of the entire system.  We present numerical results for the system sum rate of each scheme under hierarchical clustering. Our results indicate that proper design of the fog optimization results in little degradation relative to centralized optimization even for a relatively large number of virtual cells. However, improper design leads to a significant decrease in sum rate relative to centralized optimization.
\end{abstract}

%\begin{IEEEkeywords}
%component, formatting, style, styling, insert
%\end{IEEEkeywords}

\section{Introduction}
 The  demand for increased capacity in cellular networks continues to grow, and is a major driver in the deployment of 5G systems. To increase cellular network capacity, the deployment of small cells has been proposed and is currently taking place  \cite{4623708,6768783,6171992,anpalagan_bennis_vannithamby_2015}. The proximity of  small cells to one another combined with their frequency reuse can cause severe interference to neighboring small cells and macrocells, which must be managed carefully to maximize the overall network capacity. Thus, powerful interference mitigation methods as well as optimal resource allocation schemes must be developed for 5G  networks. In this work we investigate a flexible resource allocation structure for cellular systems where instead of each base-station serving all users within its own cell independently, several base-stations act cooperatively to create a “virtual cell” with joint resource allocation.

In order to design wireless networks that are composed of virtual cells we address in this work the following two design challenges: 1) Creating the virtual cells, i.e., cluster the base-stations into virtual cells. 2) Allocating the resources in each virtual cell.
This  resource allocation problem  is a non-convex NP hard problem.
In this work we address the uplink resource allocation problem for joint channel access and power allocation when there is only a single channel available  in the system. The algorithms and fundamental ideas that we present can be extended to  multi-channel systems as well as systems with interference cancellation, which is the subject of our current work.

%Recently a new concept called Software Defined Networks (SDN) was introduced \cite{6994333,6739370,7000974,1237143,7473831}. The underlying idea behind SDN is the separation of the data plane which carries the data in the network, and the control plane which determines how packets in the network are being forwarded. Theoretically, the concept of SDN can be harnessed in limiting the interference in the network by allocating the resources in the network centrally \cite{6385040,6385039}. However, due to complexity issues, the very thing that makes SDN’s centralized control plane attractive renders its implementation complexity impractical. Moreover, complexity issues are more severe in wireless communication networks because of their varying nature which requires fast updating rules for the control plane. Creating virtual cells which are composed of several cells can close the gap between the promising concept of SDN and the difficulties that arise in its implementation within wireless networks.

Base-station and user clustering as part of a resource allocation strategy is discussed in the Cooperative Multi-Point (CoMP) literature, see for example  \cite{7839266,6530435,6707857,6181826,6555174,5594575,5502468,6655533,4533793,5285181,6786390,8260866}. The work \cite{7839266} presents an extensive literature survey of cell clustering for CoMP in wireless networks. The clustering of base-stations and users can be divided into three groups: 1) Static clustering which considers a cellular network whose cells are clustered statically. Hence, the clustering does not adapt to network changes. Examples for static clustering algorithms are presented in \cite{6530435,6181826,6707857,6555174}. 2) Semi-dynamic clustering, in which static clusters are formed but the cluster affiliation of users is adapted according to the networks changes. Examples for such algorithms are presented in  \cite{5594575,5502468,6655533}. 3) Dynamic clustering in which the clustering of both base-stations and users adapts to changes in the network. Examples for dynamic clustering algorithms are presented in \cite{4533793,5285181,6786390}. 

Cell clustering as part of a resource allocation strategy in wireless networks is also investigated in the ultra-dense networks literature, see for example \cite{7008373,Tang2015,7498053,7794900,8284755,5963458}.
In particular, the work \cite{7008373} considers K-means clustering of femto access points, where each access point serves one  user. It then develops channel allocation schemes to maximize the sum rate of the downlink transmission, assuming constant power in every subchannel. The work \cite{Tang2015}  considers the downlink transmission in which users whose transmissions strongly interfere with one another are  clustered. Then, assuming that there is small number of cells in each cluster, to maximize the throughput of the network, intra-cluster interference mitigation is performed based on inter-cell interference coordination.  The work \cite{7498053} also considers downlink transmission, where sets of base-stations and users are paired, and these pairs are partitioned into clusters. The interference in each cluster is then canceled  using interference alignment in clusters; however, this limits the maximal cluster size. The work \cite{7794900} considers hierarchical clustering of cells with a non cooperative game between clusters in which each cluster aims to maximize its throughput. It assumes perfect CSI of all the channels in the network and a fixed power  allocation. The work \cite{8284755} creates clusters by maximizing the inter-cluster interference, it then allocates a different channel to each  of the clusters.  while these works address the clustering problem, they do not investigate the relation between the network clustering and resource allocation optimization scheme. Additionally, the effect of the number of clusters on the network sum rate was not analyzed.  While the aforementioned  works address the clustering problem for improving the performance of communication networks, they do not investigate the relation between the network clustering and resource allocation optimization scheme. Additionally, the effect of the number of clusters on the network sum rate is not analyzed. 
%Another relevant work is  \cite{8094947} which considers fronthaul aware resource allocation and user scheduling, however,  their methods rely on playing a stochastic game among the base-stations, assuming each has a random state, to maximize the expected sum rate. 

The remainder of this paper is organized as follows. Section \ref{sec:problem_formualtion} presents the problem formulation that we analyze in this work. Sections \ref{sec:joint_power_allocation}-\ref{sec:alternating_power_allocation_BS} present several algorithms for joint channel access and power allocation within virtual cells.
In particular, Section \ref{sec:joint_power_allocation} proposes a  joint channel access and power allocation scheme. Section  \ref{sec:alternating_power_allocation_user} proposes a channel access and power allocation scheme based on an alternating user-centric optimization.  Section
\ref{sec:alternating_power_allocation_BS} proposes a channel access and power allocation scheme based on an alternating BS-centric optimization. Section \ref{sec:virtual_cell_create} describes the method for forming the virtual cells in an optimal manner. Section \ref{se:simulation} presents numerical results of the average system sum rate for all of our proposed methods. Finally, \ref{sec:conclusion} summarize our results and discusses future work.

\section{Problem Formulation}\label{sec:problem_formualtion}
We consider  a communication network that comprises a set of base-stations (BSs) $\mathcal{B}$ and a set of users $\mathcal{U}$. The users communicate with the BSs and interfere with the transmission of one another. Each user $u\in\mathcal{U}$ has a power constraint of $\overline{P}_u$ dBm.
The BSs and users are clustered into virtual cells which must fulfill the following characteristics.

\subsection{Virtual Cells}\label{sec:virtual_cell_requirements}

\begin{definition}[Virtual BS]
Let $b_1,..,b_n$ be $n$ BSs in a communication network, we call the set $\{b_1,..,b_n\}$ a virtual BS.
\end{definition}
\begin{definition}[Proper clustering]	
Let $\mathcal{B}$ be a set of BSs,  $\mathcal{U}$ be a set of users. Denote   $\mathcal{V}=\{1,\ldots,V\}$.
For every $v$, define the sets $\mathcal{B}_v\subset \mathcal{B}$ and $\mathcal{U}_v\subset \mathcal{U}$ .
We say that the set $\mathcal{V}$ is a proper clustering of the sets  $\mathcal{B}$ and  $\mathcal{U}$  if $\mathcal{B}_v$ is a partition of the sets $\mathcal{B}$ and $\mathcal{U}$. That is,
$\bigcup_{v\in\mathcal{V}}\mathcal{B}_v = \mathcal{B}$, $\bigcup_{v\in\mathcal{U}}\mathcal{U}_v = \mathcal{U}$. Additionally,
  $\mathcal{B}_{v_1}\cap\mathcal{B}_{v_2}=\emptyset$ and $\mathcal{U}_{v_1}\cap\mathcal{U}_{v_2}=\emptyset$ for all $v_1,v_2\in\mathcal{V}$ such that $v_1\neq v_2$.
\end{definition}

\begin{definition}[Virtual cell]
Let $\mathcal{B}$ be a set of BSs, $\mathcal{U}$ be a set of users, and  $\mathcal{V}$ be a proper clustering of $\mathcal{B}$ and $\mathcal{U}$. For every $v\in\mathcal{V}$  the virtual cell  $\mathcal{C}_v$ is composed of the virtual BS $\mathcal{B}_v$ and the set of users $\mathcal{U}_v$.	
\end{definition}

This condition ensures that every BS and every user belongs to exactly one virtual cell.

Let $\mathcal{V}$ be a proper clustering of the set of BSs $\mathcal{B}$ and the set of users $\mathcal{U}$, and let   $\{\mathcal{C}_v\}_{v\in\mathcal{V}}$ be the set of virtual cells that $\mathcal{V}$ creates.
In each virtual $\mathcal{C}_v$ we assume that the BSs that compose the virtual BS $\mathcal{B}_v$ allocate their resources jointly.

\subsection{The Uplink Resource Allocation Problem}

In each virtual cell we consider the uplink resource allocation problem in which  all the BSs in the virtual cell jointly optimize the channel access and power  of the users within the virtual cell. Further, we  consider  single user detection in which every BS $b$ decodes each of its codewords separately.
That is, suppose that users $u_1$ and $u_2$ are both served by BS $b$, then $b$ decodes the codeword of $u_1$ treating the codeword of $u_2$ as noise, and decodes the codeword of $u_2$ treating the codeword of $u_1$ as noise.

While each user can communicate with all the BSs in its virtual cell, it follows by \cite{1237143} that given a power allocation scheme, the maximal communication rate for each user is achieved when the message is decoded by the BS with the highest SINR for this user.

Denote by $h_{u,b}$  the channel coefficient of the channel from user $u\in\mathcal{U}$ to BS $b$, and let $P_{u}$ be the transmit power of user $u$. Further, let $\sigma^2_{b}$ denote the noise power at BS $b$ and let $W$ denote the bandwidth of the channel.
The uplink resource allocation problem in each virtual cell $\mathcal{C}_v$, ignoring  interference from other virtual cells, is given by:
\begin{flalign}\label{eq:no_decoding_cooperation_single_discrete}
\max & \sum_{b\in\mathcal{B}_v}\sum_{u\in\mathcal{U}_v}
\gamma_{u,b}W\log_2\left(1+\frac{|h_{u,b}|^2P_{u}}{\sigma^2_{b}+J_{u,b}}\right)\nonumber\\
\text{s.t.: } &  0\leq P_{u}\leq \overline{P}_u,\quad \forall\: u\in \mathcal{U}_v,\nonumber\\
%&  \sum_{u\in\mathcal{U}_v} g_{u,\tilde{b}}P_{u}\leq \ J_{v,\tilde{b}},\quad \forall \: \tilde{b}\notin v,\nonumber\\
& \sum_{\tilde{u}\in\mathcal{U}_v ,\tilde{u}\neq u} |h_{\tilde{u},b}|^2P_{\tilde{u}}=  J_{u,b},\quad \forall \: u\in\mathcal{U}_v,b\in \mathcal{B}_v, \nonumber\\
& \sum_{b\in\mathcal{B}_v}\gamma_{u,b}\leq 1 ,\quad \forall\:u\in \mathcal{U}_v,\nonumber\\
& \gamma_{u,b}\in\{0,1\},\quad \forall\:u\in \mathcal{U}_v, b\in\mathcal{B}_v.
\end{flalign}
This is a mixed-integer programming problem that is NP-hard. The next three sections  present, respectively, three different methods to approximate the solution of this problem for a given virtual cell. The first method translates this problem from a mixed-integer programming problem to an equivalent problem with continuous variables. The second  method
approximates the optimal solution by  solving a user-centric channel access problem, and a power allocation problem, alternately.  The third  method
approximates the optimal solution by  solving a BS-centric channel access problem, and a power allocation problem, alternately.
The algorithm to define the virtual cells from all base-stations and users in the system that satisfies the conditions defined in Section \ref{sec:virtual_cell_requirements} is described in Section \ref{sec:virtual_cell_create}.

\section{Joint Channel Access and Power  Allocation}\label{sec:joint_power_allocation}
This section introduces the first  resource allocation scheme that we present in this paper. This scheme solves the power allocation and channel access problem
jointly within a given virtual cell and is found by converting (\ref{eq:no_decoding_cooperation_single_discrete}) to an equivalent continuous variable problem.
\subsection{An Equivalent Continuous Variable Resource Allocation Problem}
We can represent the problem (\ref{eq:no_decoding_cooperation_single_discrete}) by an equivalent problem with continuous variables. Suppose that instead of sending a message to the best BS a user sends messages to all BSs. The signal of user $u\in\mathcal{U}_v$ is then given by $X_u=\sum_{b\in\mathcal{B}_v}{X_{u,b}}$ where $X_{u,b}$ is the part of the signal of user $u$ intended to be decoded by BS $b$. Let $P_{u,b}$ be the power allocation of the part of the signal  of user $u$ intended to be decoded by BS $b$; i.e. $P_{u,b}=E X_{u,b}^2$.
We will argue that (\ref{eq:no_decoding_cooperation_single_discrete}) can then be written in the following equivalent form:
\begin{flalign}\label{eq:no_decoding_cooperation_single_continuous}
\max & \sum_{b\in\mathcal{B}_v}\sum_{u\in\mathcal{U}_v}
W\log_2\left(1+\frac{|h_{u,b}|^2P_{u,b}}{\sigma^2_{b}+J_{u,b}}\right)\nonumber\\
\text{s.t.: } &   \sum_{b\in\mathcal{B}_v} P_{u,b}\leq \overline{P}_u,\quad \forall\: u\in \mathcal{U}_v,\nonumber\\
%&  \sum_{u\in\mathcal{U}_v} g_{u,\tilde{b}}P_{u}\leq \ J_{v,\tilde{b}},\quad \forall \: \tilde{b}\notin v,\nonumber\\
& \sum_{\substack{(\tilde{u},\tilde{b})\in\mathcal{U}_v\times \mathcal{B}_v,\\(\tilde{u},\tilde{b})\neq (u,b)}} |h_{\tilde{u},b}|^2P_{\tilde{u},\tilde{b}}=  J_{u,b},\quad \forall \: u\in\mathcal{U}_v,b\in \mathcal{B}_v\nonumber\\
&0\leq P_{u,b},\quad \forall \: u\in\mathcal{U}_v,b\in\mathcal{B}_v.
\end{flalign}

The equivalence of (\ref{eq:no_decoding_cooperation_single_discrete}) and (\ref{eq:no_decoding_cooperation_single_continuous}) is argued as follows.
First, the solution of (\ref{eq:no_decoding_cooperation_single_discrete}) can be achieved by the solution of (\ref{eq:no_decoding_cooperation_single_continuous}) by setting $X_{u,b}=0$ whenever $\gamma_{u,b}=0$, and $E X_{u,b}^2 = P_u$ whenever $\gamma_{u,b}=1$; thus thus maximal sum rate  (\ref{eq:no_decoding_cooperation_single_continuous}) upper bounds the maximal sum rate  (\ref{eq:no_decoding_cooperation_single_discrete}). On the other hand,  it follows by \cite{1237143} that the maximal sum rate of (\ref{eq:no_decoding_cooperation_single_continuous}) cannot be larger than that of (\ref{eq:no_decoding_cooperation_single_discrete}). Thus, the two problems (\ref{eq:no_decoding_cooperation_single_discrete}) and (\ref{eq:no_decoding_cooperation_single_continuous}) are equivalent.

\subsection{Solving an Approximation of the Continuous Variable Resource Allocation Problem}
Denote:
\begin{flalign}
\text{SINR}_{u,b}(\boldsymbol P) =\frac{|h_{u,b}|^2 P_{u,b}}{\sigma^2_b+\sum_{\substack{(\tilde{u},\tilde{b})\in\mathcal{U}_v\times \mathcal{B}_v,\\(\tilde{u},\tilde{b})\neq (u,b)}} |h_{\tilde{u},\tilde{b}}|^2P_{\tilde{u},\tilde{b}}},
\end{flalign}
where $\boldsymbol P = (P_{u,b})_{(u,b)\in\mathcal{U}_{v}\times\mathcal{B}_{v}}$ is the matrix of the transmission power.

Using the high SINR approximation \cite{5165179}
\begin{flalign}\label{eq:high_SINR_approx_improved}
\log(1+z)\geq \alpha(z_0)\log z+\beta(z_0),
\end{flalign}
where
\begin{flalign}\label{eq:alpha_beta_def}
\alpha(z_0) &= \frac{z_0}{1+z_0},\quad\beta(z_0) =\log(1+z_0)-\frac{z_0}{1+z_0}\log{z_0},
\end{flalign}
yields the following approximated iterative problem:

\begin{flalign}\label{eq:iterative_alpha_approx}
&\boldsymbol{P}^{(m)} = \arg\max_{\boldsymbol P}  \sum_{b\in\mathcal{B}_v}\sum_{u\in\mathcal{U}_v}
W\cdot\nonumber\\
&\hspace{0.7cm}\left[\alpha_{u,b}^{(m)}\log_2\left(\frac{|h_{u,b}|^2P_{u,b}}{\sigma^2_{b}+\sum_{\substack{(\tilde{u},\tilde{b})\in\mathcal{U}_v\times \mathcal{B}_v,\\(\tilde{u},\tilde{b})\neq (u,b)}} |h_{\tilde{u},b}|^2P_{\tilde{u},\tilde{b}}}\right)
+\beta_{u,b}^{(m)}\right]\nonumber\\
&\text{s.t.: }    \sum_{b\in\mathcal{B}_v} P_{u,b}\leq \overline{P}_u,\quad 0\leq P_{u,b},\quad \forall \: u\in\mathcal{U}_v,b\in\mathcal{B}_v,
\end{flalign}

where $\alpha_{u,b}^{(m)}=\alpha(\text{SINR}_{u,b}(\boldsymbol P^{(m-1)}))$, $\beta_{u,b}^{(m)}=\beta(\text{SINR}_{u,b}(\boldsymbol P^{(m-1)}))$ and $\alpha_{u,b}^{(0)}=1$, $\beta_{u,b}^{(0)}=0$ for all $u\in\mathcal{U}_v$ and $b\in\mathcal{B}_v$.
It is left to solve the problem (\ref{eq:iterative_alpha_approx}). Transforming the variables of the problem using $P_{u,b}=\exp(g_{u,b})$  yields the equivalent convex problem:
\begin{flalign}\label{sol_continuous_power_approx}
&\ln(\boldsymbol P^{(m)}) = \arg\max \sum_{b\in\mathcal{B}_v}\sum_{u\in\mathcal{U}_v}
W\alpha_{u,b}^{(m)}\left[\frac{g_{u,b}}{\ln2}+\log_2(|h_{u,b}|^2)\vphantom{\log_2\left(\sigma^2_{b}+\sum_{\substack{(\tilde{u},\tilde{b})\in\mathcal{U}_v\times \mathcal{B}_v,\\(\tilde{u},\tilde{b})\neq (u,b)}} |h_{\tilde{u},b}|^2\exp(g_{\tilde{u},\tilde{b}})\right)}\right.\nonumber\\
&\hspace{1.8cm}\left.-\log_2\left(\sigma^2_{b}+\sum_{\substack{(\tilde{u},\tilde{b})\in\mathcal{U}_v\times \mathcal{B}_v,\\(\tilde{u},\tilde{b})\neq (u,b)}} |h_{\tilde{u},b}|^2\exp(g_{\tilde{u},\tilde{b}})\right)\right]\nonumber\\
&\text{s.t.: }    \sum_{b\in\mathcal{B}_v} \exp(g_{u,b})\leq \overline{P}_u,\quad \forall\: u\in \mathcal{U}_v.
\end{flalign}
Since the problem (\ref{sol_continuous_power_approx}) is convex with nonempty interior, its duality gap is zero. Define 
\begin{flalign*}%\label{update_rule_continuous}
&f_{u,b}(\boldsymbol{P},\lambda,\boldsymbol\alpha)=\nonumber\\
&\hspace{0.7cm}\frac{W\alpha_{u,b}}{\lambda\ln 2+W\sum_{\substack{(\tilde{u},\tilde{b})\in\mathcal{U}_v\times \mathcal{B}_v,\\(\tilde{u},\tilde{b})\neq (u,b)}}\alpha_{\tilde{u},\tilde{b}}\frac{\text{SINR}_{\tilde{u},\tilde{b}}(\boldsymbol P)}{P_{\tilde{u},\tilde{b}}|h_{\tilde{u},\tilde{b}}|^2 }|h_{u,\tilde{b}}|^2}.
\end{flalign*}
Let $\lambda_u$ be the Lagrangian multiplier associated with the power constraint of user $u$ in (\ref{sol_continuous_power_approx}).
By \cite{5165179} we can find the power allocation that achieves the Lagrangian dual function of (\ref{sol_continuous_power_approx}) by the following  fixed point iteration:
\begin{flalign*}
&P_{u,b}^{(m,s+1)}=f_{u,b}(\boldsymbol{P}^{(m,s)},\lambda_u,\boldsymbol\alpha^{(m)}).
\end{flalign*}
We can then solve the dual problem of (\ref{sol_continuous_power_approx}) by  optimizing the Lagrangian dual function over $\lambda_u$ using gradient methods.
Alternatively, by the convexity of (\ref{sol_continuous_power_approx}), the following KKT conditions are sufficient for optimality:
\begin{flalign*}
&P_{u,b}^{(m)}=f_{u,b}(\boldsymbol{P}^{(m)},\lambda_u,\boldsymbol\alpha^{(m)}),\nonumber\\
&0=\lambda_u\left(\sum_{b\in\mathcal{B}_v} P_{u,b}^{(m)}- \overline{P}_u\right),\quad \sum_{b\in\mathcal{B}_v} P_{u,b}^{(m)}\leq \overline{P}_u,\quad\lambda_u\geq 0,
\end{flalign*}
for all $u\in\mathcal{U}_v$.
Define the following fixed point iteration:
\begin{flalign}\label{update_rule_continuous}
&P_{u,b}^{(m,s+1)}=f_{u,b}(\boldsymbol{P}^{(m,s)},\lambda_u^{(s)},\boldsymbol\alpha^{(m)}),
\end{flalign}
where $\lambda_u^{(s)}=0$ if
\begin{flalign}
\sum_{b\in\mathcal{B}_v}f_{u,b}(\boldsymbol{P}^{(m,s)},0,\boldsymbol\alpha^{(m)})\leq\overline{P}_u,
\end{flalign}
otherwise $\lambda_u^{(s)}$ is chosen such that $\sum_{b\in\mathcal{B}_v}P_{u,b}^{(m,s+1)}=\overline{P}_u$.
While the conditions for the convergence of (\ref{update_rule_continuous}) presented in \cite{414651} are not fulfilled, in practice the convergence of (\ref{update_rule_continuous}) is observed. Note that whenever (\ref{update_rule_continuous}) converges, it converges to an optimal point of (\ref{sol_continuous_power_approx}).

\section{User Centric Resource Allocation}\label{sec:alternating_power_allocation_user}
This section presents the second  resource allocation scheme that we consider. This scheme is composed of an alternating algorithm for the resource allocation problem (\ref{eq:no_decoding_cooperation_single_discrete}). This algorithm is  user-centric in that it starts from a maximal power allocation in which every user transmits with its maximal power. Then,  every user chooses the receiving BS to be the one  with the maximal SINR for this user. Then, we alternate between the power allocation and channel access problems as described in Algorithm \ref{algo:Antenating_single_set_power} below.

\begin{algorithm}
	\caption{}\label{algo:Antenating_single_set_power}
	\begin{algorithmic}[1]
		%\Procedure{MyProcedure}{}
		\State Input: $\delta>0$
		\State Set $n=0$,  $\delta_0 = 2\delta$, $R_0 = 0$;
		\State Set $P^{(1)}_u=\overline{P}_u$ for all $u\in\mathcal{U}_v$;
		\While{  $\delta_n>\delta$}
		\State $n=n+1$;
		\State For every $u\in\mathcal{U}_v$ and $b\in\mathcal{B}_v$ calculate
		\[ J^{(n)}_{u,b}=\sum_{\tilde{u}\in\mathcal{U}_v ,\tilde{u}\neq u} |h_{\tilde{u},b}|^2P_{\tilde{u}}^{(n)} ,\quad \forall \: u\in\mathcal{U}_v,b\in \mathcal{B}_v;\]
		\State For every $u\in\mathcal{U}_v$, calculate \[b_u^{(n)} = \arg\max_{b\in\mathcal{B}_v} \frac{|h_{u,b}|^2P^{(n)}_{u}}{\sigma^2_{b}+J^{(n)}_{u,b}};\]
		\State For every $u\in\mathcal{U}_v$ and $b\in\mathcal{B}_v$  set $\gamma^{(n)}_{u,b}=\mathbbm{1}_{\{b = b^{(n)}_u\}}$;
		\State Calculate the sum rate \[R_n =\sum_{b\in\mathcal{B}_v}\sum_{u\in\mathcal{U}_v}
		\gamma^{(n)}_{u,b}W\log_2\left(1+\frac{|h_{u,b}|^2P^{(n)}_{u}}{\sigma^2_{b}+J^{(n)}_{u,b}}\right); \]
		\State $\delta_n = R_n-R_{n-1}$;
		\State Calculate the optimal power allocation $(P_u^{(n+1)})_{u\in\mathcal{U}_v}$ given the channel allocation $(\gamma_{u,b}^{(n)})_{(u,b)\in\mathcal{U}_v\times\mathcal{B}_v}$ by solving:
		\begin{flalign}\label{eq:power_alloc_algo1}
		&(P_u^{(n+1)})_{u\in\mathcal{U}_v}=\nonumber\\
		&\arg\max_{P_{u}}  \sum_{b\in\mathcal{B}_v}\sum_{u\in\mathcal{U}_v}
		\gamma_{u,b}^{(n)}W\log_2\left(1+\frac{|h_{u,b}|^2P_{u}}{\sigma^2_{b}+J_{u,b}}\right)\nonumber\\
		&\text{s.t.: }   0\leq P_{u}\leq \overline{P}_u,\quad \forall\: u\in \mathcal{U}_v,\nonumber\\
		%&  \sum_{u\in\mathcal{U}_v} g_{u,\tilde{b}}P_{u}\leq \ J_{v,\tilde{b}},\quad \forall \: \tilde{b}\notin v,\nonumber\\
		& \sum_{\tilde{u}\in\mathcal{U}_v ,\tilde{u}\neq u} |h_{\tilde{u},b}|^2P_{\tilde{u}}=  J_{u,b},\quad \forall \: u\in\mathcal{U}_v,b\in \mathcal{B}_v;
		\end{flalign}
		
		\EndWhile
		%\EndProcedure
	\end{algorithmic}
	
\end{algorithm}

It is left to solve the  problem (\ref{eq:power_alloc_algo1}).
First, we rewrite this problem by defining the function $b(u):\mathcal{U}\rightarrow\mathcal{B}$ to be the BS that decodes the message of user $u$, it follows that $\gamma_{u,b(u)}=1$.
The optimization problem  (\ref{eq:power_alloc_algo1}) can then be written as,
\begin{flalign}\label{eq:power_alloc_algo1_equi}
\max&\sum_{u\in\mathcal{U}_v}
W\log_2\left(1+\frac{|h_{u,b(u)}|^2P_{u}}{\sigma^2_{b(u)}+\sum_{\tilde{u}\in\mathcal{U}_v ,\tilde{u}\neq u} |h_{\tilde{u},b(u)}|^2P_{\tilde{u}}}\right)\nonumber\\
\text{s.t.: } &  0\leq P_{u}\leq \overline{P}_u,\quad \forall\: u\in \mathcal{U}_v,
%&  \sum_{u\in\mathcal{U}_v} g_{u,\tilde{b}}P_{u}\leq \ J_{v,\tilde{b}},\quad \forall \: \tilde{b}\notin v.
\end{flalign}

Denote:
\begin{flalign}
\text{SINR}_{u,b(u)}(\boldsymbol P) = \frac{|h_{u,b(u)}|^2P_{u}}{\sigma^2_{b(u)}+\sum_{\tilde{u}\in\mathcal{U}_v ,\tilde{u}\neq u} |h_{\tilde{u},b(u)}|^2P_{\tilde{u}}},
\end{flalign}
where $\boldsymbol{P} = (P_u)_{u\in\mathcal{U}}$ is the vector of the power transmission.

We solve the problem (\ref{eq:power_alloc_algo1_equi}) approximately by applying the high SINR approximation (\ref{eq:high_SINR_approx_improved}). This yields the problem

\begin{flalign}\label{sol_alter_power_approx_user}
&\boldsymbol{P}^{(m)} = \arg\max_{\boldsymbol P} \sum_{u\in\mathcal{U}_v}
W\cdot\nonumber\\
&\hspace{0.7cm}\left[\alpha_{u}^{(m)}\log_2\left(\frac{|h_{u,b(u)}|^2P_{u}}{\sigma^2_{b(u)}+\sum_{\tilde{u}\in\mathcal{U}_v ,\tilde{u}\neq u} |h_{\tilde{u},b(u)}|^2P_{\tilde{u}}}\right)+\beta_u^{(m)}\right]\nonumber\\
&\text{s.t.: }  0\leq P_{u}\leq \overline{P}_u,\quad \forall\: u\in \mathcal{U}_v,
%&  \sum_{u\in\mathcal{U}_v} g_{u,\tilde{b}}P_{u}\leq \ J_{v,\tilde{b}},\quad \forall \: \tilde{b}\notin v,
\end{flalign}
where $\alpha_u^{(m)} = \alpha(\text{SINR}_{u,b(u)}(\boldsymbol P^{(m-1)}))$ and $\beta_u^{(m)} = \beta(\text{SINR}_{u,b(u)}(\boldsymbol P^{(m-1)}))$, and $\alpha_u^{(0)}=1$, $\beta_u^{(0)}=0$ for all $u\in\mathcal{U}_v$.

Now,  substituting $P_u=\exp(g_u)$  yields the equivalent problem
\begin{flalign}\label{sol_alter_power_approx_user_ln}
&\ln(\boldsymbol{P}^{(m)}) = \arg\max \sum_{u\in\mathcal{U}_v}
W\alpha_u^{(m)}\cdot\nonumber\\
&\hspace{2.1cm}\log_2\left(\frac{|h_{u,b(u)}|^2\exp(g_u)}{\sigma^2_{b(u)}+\sum_{\tilde{u}\in\mathcal{U}_v ,\tilde{u}\neq u} |h_{\tilde{u},b(u)}|^2\exp(g_{\tilde u})}\right)\nonumber\\
&\text{s.t.: }   \exp(g_u)\leq \overline{P}_u,\quad \forall\: u\in \mathcal{U}_v.
%&  \sum_{u\in\mathcal{U}_v} g_{u,\tilde{b}}P_{u}\leq \ J_{v,\tilde{b}},\quad \forall \: \tilde{b}\notin v,
\end{flalign}

Since the problem (\ref{sol_alter_power_approx_user_ln}) is convex with non empty interior, its duality gap is zero and the KKT conditions are sufficient for the points to be primal and dual optimal. The KKT conditions for (\ref{sol_alter_power_approx_user_ln}), after substituting $P_u=\exp(g_u)$, are
\begin{flalign}
&P_u^{(m)}=\frac{W\alpha_u^{(m)}}{\lambda_u\ln 2+W\sum_{\substack{\tilde{u}\in\mathcal{U}_v,\\ \tilde{u}\neq u}}\frac{\alpha_{\tilde{u}}^{(m)}|h_{u,b(\tilde{u})}|^2}{\sigma^2_{b(\tilde{u})}+\sum_{\substack{\hat{u}\in\mathcal{U}_v ,\\\hat{u}\neq \tilde{u}}} |h_{\hat{u},b(\tilde{u})}|^2P_{\hat u}^{(m)}}}\nonumber\\
&0=\lambda_u\left( P_u^{(m)}- \overline{P}_u\right),\quad P_u^{(m)}\leq \overline{P}_u,\quad\lambda_u\geq 0,
\end{flalign}
for all $u\in\mathcal{U}_v$.
Thus, the power allocation for user $u$ must fulfill the equation:
\begin{flalign}\label{eq:Pu_assigment}
P_u^{(m)}=\min\left\{\overline{P}_u,\frac{\alpha_u^{(m)}}{\sum_{\substack{\tilde{u}\in\mathcal{U}_v,\\ \tilde{u}\neq u}}\alpha_{\tilde{u}}^{(m)}\frac{\text{SINR}_{\tilde{u},b(\tilde{u})}(\boldsymbol P^{(m)})}{P_{\tilde{u}}^{(m)}|h_{\tilde{u},b(\tilde{u})}|^2 }|h_{u,b(\tilde{u})}|^2}\right\}.
\end{flalign}
By \cite{414651} we have that we can solve (\ref{eq:Pu_assigment}) iteratively. Let $P_u^{(m,s)}$ be the transmitting power of user $u$ at the $s$th iteration.  Then the update rule
\begin{flalign}\label{eq:Pu_assigment_elegant}
&P_u^{(m,s+1)}=\nonumber\\
&\min\left\{\overline{P}_u,\frac{\alpha_u^{(m)}}{\sum_{\substack{\tilde{u}\in\mathcal{U}_v,\\ \tilde{u}\neq u}}\alpha_{\tilde{u}}^{(m)}\frac{\text{SINR}_{\tilde{u},b(\tilde{u})}(\boldsymbol P^{(m,s)})}{P_{\tilde{u}}^{(m,s)}|h_{\tilde{u},b(\tilde{u})}|^2 }|h_{u,b(\tilde{u})}|^2}\right\}
\end{flalign}
converges to the solution of (\ref{eq:Pu_assigment}), both synchronously and asynchronously, provided that it exists. The existence of the solution is guaranteed because of the strong convexity of (\ref{sol_alter_power_approx_user_ln}).

We note that we can also solve (\ref{sol_alter_power_approx_user}) by solving the problem (\ref{eq:iterative_alpha_approx}) using the initial values:
$\alpha_{u,b}^{(0)} = \mathbbm{1}_{\{b=b(u)\}}$ and $\beta_{u,b}^{(0)}=0$ for all $(u,b)\in\mathcal{U}_v\times\mathcal{B}_v$.

\section{Base-Station Centric Resource Allocation}\label{sec:alternating_power_allocation_BS}
This section presents the third and final resource allocation scheme that is included in this paper. This scheme is composed of an alternating algorithm for the resource allocation problem (\ref{eq:no_decoding_cooperation_single_discrete}). This algorithm optimizes the resource allocation in a BS centric manner. In particular it starts from a maximal power allocation in which every user transmits with its maximal power. Then,  every BS chooses the transmitting user to be the one  with the maximal SINR for this BS. Then, we alternate between the power allocation and channel access problems as described in Algorithm \ref{algo:Antenating_single_set_power_BS}.
\newline

\begin{algorithm}
	\caption{}\label{algo:Antenating_single_set_power_BS}
	\begin{algorithmic}[1]
		%\Procedure{MyProcedure}{}
		\State Input: $\delta>0$
		\State Set $n=0$,  $\delta_0 = 2\delta$, $R_0 = 0$;
		\State Set $P^{(1)}_u=\overline{P}_u$ for all $u\in\mathcal{U}_v$;
		\While{  $\delta_n>\delta$}
		\State $n=n+1$;
		\State For every $b\in\mathcal{B}_v$ and $u\in\mathcal{U}_v$ calculate
		\[ J^{(n)}_{u,b}=\sum_{\tilde{u}\in\mathcal{U}_v ,\tilde{u}\neq u} |h_{\tilde{u},b}|^2P_{\tilde{u}}^{(n)} ,\quad \forall \: u\in\mathcal{U}_v,b\in \mathcal{B}_v;\]
		\State For every $u\in\mathcal{U}_v$, calculate \[u_b^{(n)} = \arg\max_{u\in\mathcal{U}_v} \frac{|h_{u,b}|^2P^{(n)}_{u}}{\sigma^2_{b}+J^{(n)}_{u,b}};\]
		\State For every $b\in\mathcal{B}_v$ and $u\in\mathcal{U}_v$  set $\gamma^{(n)}_{u,b}=\mathbbm{1}_{\{u = u^{(n)}_b\}}$;
		\State Calculate the sum rate \[R_n =\sum_{b\in\mathcal{B}_v}\sum_{u\in\mathcal{U}_v}
		\gamma^{(n)}_{u,b}W\log_2\left(1+\frac{|h_{u,b}|^2P^{(n)}_{u}}{\sigma^2_{b}+J^{(n)}_{u,b}}\right); \]
		\State $\delta_n = R_n-R_{n-1}$;
		\State Calculate the optimal power allocation $(P_u^{(n+1)})_{u\in\mathcal{U}_v}$ given the channel allocation $(\gamma_{u,b}^{(n)})_{(u,b)\in\mathcal{U}_v\times\mathcal{B}_v}$ by solving:
		\begin{flalign}\label{eq:power_alloc_algo2_BS}
		& (P_{u,b}^{(n+1)})_{u\in\mathcal{U}_v,b\in\mathcal{B}_v}=\nonumber\\
		& \arg\max  \sum_{b\in\mathcal{B}_v}\sum_{u\in\mathcal{U}_v}
		\gamma_{u,b}^{(n)}W\log_2\left(\frac{|h_{u,b}|^2P_{u,b}}{\sigma^2_{b}+J_{u,b}}\right)
		\nonumber\\
&\text{s.t.: }   0\leq \sum_{b\in\mathcal{B}_v}P_{u,b}\leq \overline{P}_u,\quad \forall\: u\in \mathcal{U}_v,\nonumber\\
& \sum_{\substack{\tilde{u}\in\mathcal{U}_v,\tilde{b}\in\mathcal{B}_v,\\ (\tilde{u},\tilde{b})\neq(u,b)} } |h_{\tilde{u},b}|^2P_{\tilde{u},\tilde{b}}=  J_{u,b},\quad \forall \: u\in\mathcal{U}_v,b\in \mathcal{B}_v;
\end{flalign}
\EndWhile
%\EndProcedure
\end{algorithmic}
\end{algorithm}

It is left to solve the  problem (\ref{eq:power_alloc_algo2_BS}) .
We rewrite this problem by defining the function $u(b):\mathcal{B}_v\rightarrow\mathcal{U}_v$ to be the user that transmit tp BS $b$, it follows that $\gamma_{u(b),b}=1$.
The optimization problem  (\ref{eq:power_alloc_algo2_BS}) is then can be written as,
\begin{flalign}\label{eq:power_alloc_algo2_equi}
\max&\sum_{b\in\mathcal{B}_v}
W\log_2\left(1+\frac{|h_{u(b),b}|^2P_{u(b),b}}{\sigma^2_{b}+\sum_{\tilde{b}\in\mathcal{B}_v, \tilde{b}\neq b}  |h_{u(\tilde{b}),b}|^2P_{u(\tilde{b}),\tilde{b}}}\right)\nonumber\\
\text{s.t.: } &  0\leq \sum_{b\in\mathcal{B}_v:u(b)=u} P_{u,b}\leq \overline{P}_u,\quad \forall\: u\in \mathcal{U}_v,
%&  \sum_{u\in\mathcal{U}_v} g_{u,\tilde{b}}P_{u}\leq \ J_{v,\tilde{b}},\quad \forall \: \tilde{b}\notin v.
\end{flalign}

Denote:
\begin{flalign}
\text{SINR}_{u(b),b}(\boldsymbol P) = \frac{|h_{u(b),b}|^2P_{u(b),b}}{\sigma^2_{b}+\sum_{\tilde{b}\in\mathcal{B}_v, \tilde{b}\neq b}  |h_{u(\tilde{b}),b}|^2P_{u(\tilde{b}),\tilde{b}}},
\end{flalign}                                      
where $\boldsymbol{P} = (P_{u(b),b})_{b\in\mathcal{B}_v}$ is the vector of the power transmission.
We solve the problem (\ref{eq:power_alloc_algo1_equi}) approximately by applying the high SINR approximation (\ref{eq:high_SINR_approx_improved}). This yields the problem
\begin{flalign}\label{sol_alter_power_approx_BS}
&\boldsymbol{P}^{(m)} = \arg\max_{\boldsymbol P} \sum_{b\in\mathcal{B}_v}
W\cdot\nonumber\\
&\hspace{0.5cm}\left[\alpha_{b}^{(m)}\log_2\left(\frac{|h_{u(b),b}|^2P_{u(b),b}}{\sigma^2+\sum_{\tilde{b}\in\mathcal{B}_v, \tilde{b}\neq b}  |h_{u(\tilde{b}),b}|^2P_{u(\tilde{b}),\tilde{b}}}\right)+\beta_b^{(m)}\right]\nonumber\\
&\hspace{1.5cm}\text{s.t.: }   0\leq \sum_{b\in\mathcal{B}_v:u(b)=u} P_{u}\leq \overline{P}_{u,b},\quad \forall\: u\in \mathcal{U}_v,
%&  \sum_{u\in\mathcal{U}_v} g_{u,\tilde{b}}P_{u}\leq \ J_{v,\tilde{b}},\quad \forall \: \tilde{b}\notin v,
\end{flalign}                                                                         where $\alpha_b^{(m)} = \alpha(\text{SINR}_{u(b),b}(\boldsymbol P^{(m-1)}))$ and $\beta_b^{(m)} = \beta(\text{SINR}_{u(b),b}(\boldsymbol P^{(m-1)}))$, and $\alpha_b^{(0)}=1$, $\beta_b^{(0)}=0$ for all $b\in\mathcal{B}_v$.
We  solve (\ref{sol_alter_power_approx_BS}) by solving the problem (\ref{eq:iterative_alpha_approx}) using the initial values:
$\alpha_{u,b}^{(0)} = \mathbbm{1}_{\{u=u(b)\}}$ and $\beta_{u,b}^{(0)}=0$ for all $(u,b)\in\mathcal{U}_v\times\mathcal{B}_v$.

\section{Forming the Virtual Cells}\label{sec:virtual_cell_create}
This section presents the clustering approaches that create the virtual cells within which the resource allocation algorithms defined in the previous three sections operate.
We consider two methods to cluster the BSs. The first is a hierarchical clustering of the BS according to a minimax linkage criteria. To evaluate the performance of this clustering method we compare it  to an exhaustive search over all the possible clusters.
\subsection{Base-Station Clustering}
\paragraph{Hierarchical clustering - Minimax linkage \cite{BienTibshirani2011}}
Let $d:\mathbb{R}^2\times\mathbb{R}^2\rightarrow\mathbb{R}$ be the Euclidean distance function.
\begin{definition}[Radius of a set around point]
Let $S$ be a set of points in $\mathbb{R}^2$, the radius of $S$ around $s_i \in S$  is defined as $r(s_i,S)=\max_{s_j\in S}\:d(s_i,s_j)$.
\end{definition}
\begin{definition}[Minimax radius]
Let $S$ be a set of points in $\mathbb{R}^l$, the minimax radius of $S$ is defined as $r(S) = \min_{s_i\in S}\: r(s_i,S)$.
\end{definition}
\begin{definition}[Minimax linkage]
The minimax linkage between two set of points $S_1$ and $S_2$ in $\mathbb{R}^l$ is defines as $d(S_1,S_2) = r(S_1\cup S_2)$.

Note that $d(\{s_1\},\{s_2\}) = r(\{s_1\}\cup \{s_2\})=d(s_1,s_2)$.
\end{definition}

The  agglomerative hierarchical clustering algorithm using the minimax linkage criterion is as follows:

\textbf{Input:} $C_n = \left\{\{s_1\},\dots,\{s_n\}\right\}$ and $d(\{s_i\},\{s_j\})=d(s_i,s_j),\:\forall s_i,s_j\in S=\{s_1,\ldots,s_n\}$.

For all $k = n-1,\ldots,1$
\begin{enumerate}
\item Find $(S_1,S_2) = \arg\min_{G,H\in C_k } d(G,H).$
\item  Update $C_{k} = C_{k+1} \bigcup \{S_1\cup S_2\} \setminus \{S_1,S_2\}$.
\item Calculate $d(S_1\cup S_2,G)$ for all $G\in C_k$.
\end{enumerate}

We perform the hierarchical clustering over the set of BS locations to create the virtual BSs.

The hierarchical clustering is of great relevance to our setup since it enjoys an important property that both the K-means clustering and the spectral clustering lack, namely, the number of clusters can be changed without disassembling all the clusters in the networks. Thus, the  number of virtual BSs can be easily adapted according to current state of the network.

\paragraph{Exhaustive Search} As previously written, to evaluate  the performance of hierarchical clustering  we  performed exhaustive search over all the possible clustering. In this way, for each number of clusters (virtual cells) we produced all the possible clusters and in the end chose the clustering that yielded the maximal sum rate of the network, considering interference from other virtual cells,  given that particular number of clusters.

\subsection{Users' Affiliation with Clusters}\label{sec_user_affil}
To create the virtual cells, each user is affiliated with its closest BS, and belongs to the cluster its affiliated BS belongs to; this way every virtual BS and it associated users compose a virtual cell.

It is easy to verify that this formation of the virtual cells fulfills the requirement presented in Section \ref{sec:virtual_cell_requirements}.

\section{Numerical Results}\label{se:simulation}
This section presents Monte Carlo simulation results that compares the three resource allocation schemes for both the hierarchical clustering and the exhaustive search over all possible clustering. We set the following parameters for the simulation: the network is comprised of $6$ BSs and $30$ users which were uniformly located in a square of side $2500$ meter. The channel bandwidth was $1$ MHz and the carrier frequency was  $1800$ MHz. The noise power received by each BS was $-174$ dBm/Hz, and the maximal power constraint for each user was $23$ dBm. Finally, we simulated the channel gains according to a free space model $h_{u,b} = \frac{\lambda}{4\pi}\cdot\frac{1}{d(u,b)}$,  where $\lambda$ is the signal wavelength, and $d(u,b)$ is the distance between user $u$ and BS $b$.
We averaged the results over 500 measurements, in each we generated randomly the locations of the BSs and users.

In the simulation we compared the sum rate achieved by each of the clustering and resource allocation methods presented in this paper. We note that while the resource allocation ignored the interference caused by other virtual cells, the sum rate of the network was calculated considering this interference.

\begin{figure}
\centering
\includegraphics[scale=0.25]{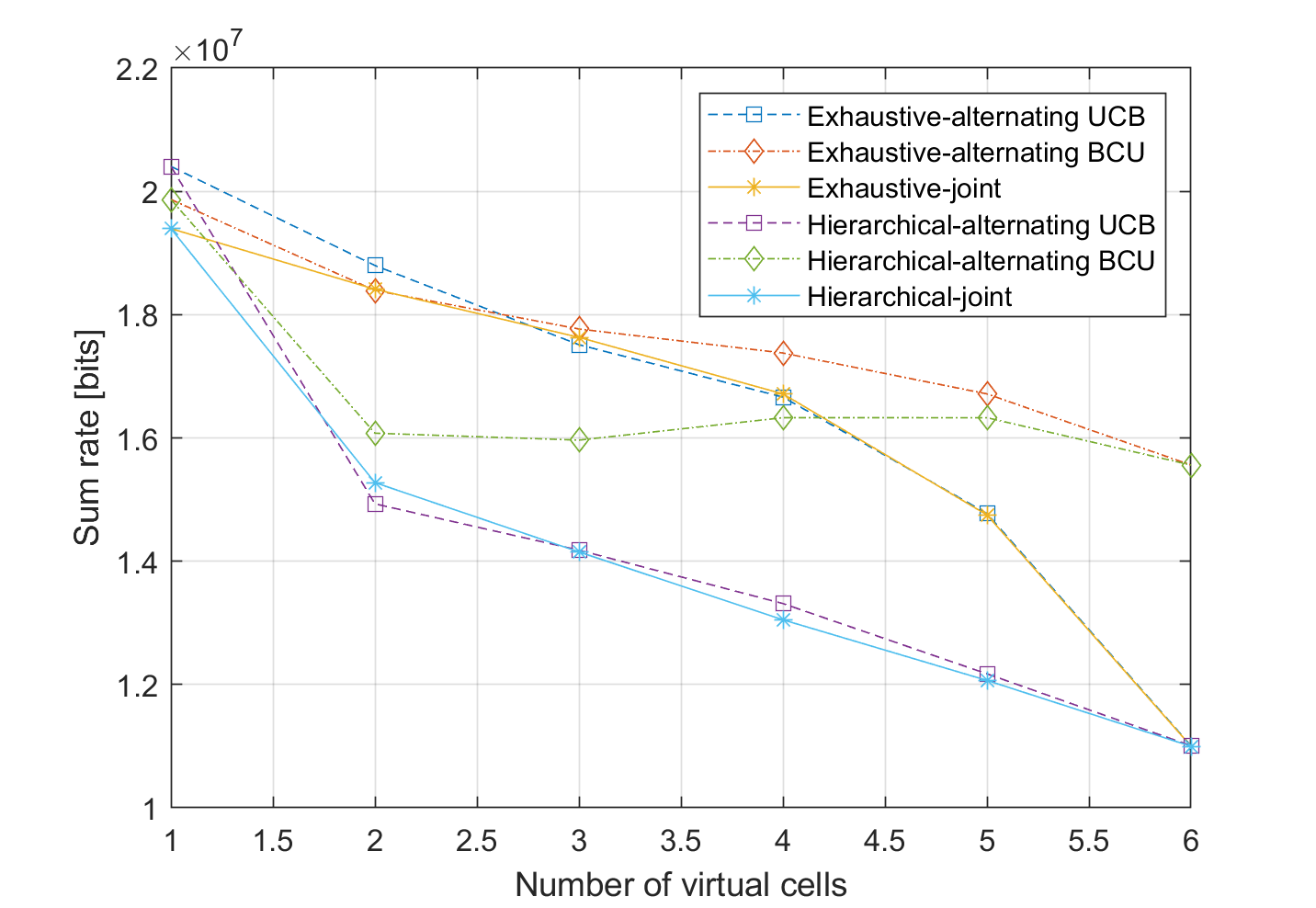}
%,clip,trim=4cm 8.5cm 4.5cm 9cm
\caption{Sum rate as a function of the number of clusters.}
\label{single_sum_rate_fig}
\end{figure}
The simulation results are shown in Fig. \ref{single_sum_rate_fig} for the following cases:
\begin{itemize}
\item Exhaustive - alternating UCB and Hierarchical - alternating UCB (user choose base-station) lines refer to the resource allocation scheme presented in Section   \ref{sec:alternating_power_allocation_user}  when performing exhaustive search over all the possible clusters, and hierarchical clustering, respectively.
\item Exhaustive - alternating BCU and Hierarchical - alternating BCU (base-station choose user) lines refer to the resource allocation scheme presented in Section \ref{sec:alternating_power_allocation_BS}  when performing exhaustive search over all the possible clusters, and hierarchical clustering, respectively..
\item Exhaustive -joint  and Hierarchical -joint lines refer to the resource allocation scheme presented in Section \ref{sec:joint_power_allocation} when performing exhaustive search over all the possible clusters, and hierarchical clustering, respectively.
%\item Hierarchical - alternating UCB (user choose base-station) line refers to the resource allocation scheme presented in Section    \ref{sec:alternating_power_allocation_user} when performing hierarchical clustering.
%\item Hierarchical - alternating BCU (base-station choose user) line refers to the resource allocation scheme presented in Section \ref{sec:alternating_power_allocation_BS} when performing performing hierarchical clustering.
%\item Hierarchical -joint line line refers to the resource allocation scheme presented in Section \ref{sec:joint_power_allocation} when performing performing hierarchical clustering.
\end{itemize}

Figure \ref{single_sum_rate_fig} leads to several interesting insights and conclusions, first, it confirms the expectation that as the number of virtual cells decreases the average sum rate increases. The only line that does not show this is the Hierarchical - alternating BCS line in which interference from outer virtual cells was more severe. For a larger number of virtual cells the BS centric resource allocation outperformed the other two methods, however, for a smaller number of virtual cells  the user centric resource allocation outperformed the other two methods.
As for the joint channel and power allocation, even though its performance did not exceed the other two, the algorithm can be improve in the future by choosing better initial values for $\alpha_{u,b}$ and $\beta_{u,b}$. The initial values of the two other resource allocation methods are two examples for such initial values.
Finally, comparing the performance of the exhaustive search over all the possible virtual cells of a certain size and the hierarchical clustering leads to the conclusion that the hierarchical clustering can improve the sum rate of the network compared to the independent resource allocation by each BS (i.e. the current fully distributed optimization method), however, this depends on the resource allocation scheme. Thus, future evaluation and the development of  resource allocation and clustering schemes must go hand in hand since they both significantly affect the performance of the network.

\section{Conclusion and Future Work}\label{sec:conclusion}
This work addresses the role of virtual cells in resource allocation for future wireless networks. It proposes methods for two design aspects of this optimization; namely, forming the virtual cells and allocating the communication resources in each virtual cell effectively. We present three types of resource allocation schemes: the first converts the mixed integer resource allocation problem into a continuous resource allocation problem and then finds an approximate solution, the second alternates between the power allocation and channel access problems when the channel allocation is carried out in a user-centric manner, finally, the third resource allocation scheme we present alternates between the power allocation and channel access problems when the channel allocation is carried out from a base-station centric perspective.  We also propose the use of  hierarchical clustering  in the clustering of the base-stations to form the virtual cells, since changing the number of virtual cells only causes local changes and does not force a recalculation of all the virtual base-stations in the network. Finally, we present numerical results for all these methods and discuss the merits of the resource allocation and clustering schemes. We note that the results presented in this paper can be extended to the multi channel setup, we also note that other  hierarchical  clustering algorithms can be considered in order to improve the overall network performance.
 
\bibliographystyle{IEEEtran}

\begin{thebibliography}{10}
	\providecommand{\url}[1]{#1}
	\csname url@samestyle\endcsname
	\providecommand{\newblock}{\relax}
	\providecommand{\bibinfo}[2]{#2}
	\providecommand{\BIBentrySTDinterwordspacing}{\spaceskip=0pt\relax}
	\providecommand{\BIBentryALTinterwordstretchfactor}{4}
	\providecommand{\BIBentryALTinterwordspacing}{\spaceskip=\fontdimen2\font plus
		\BIBentryALTinterwordstretchfactor\fontdimen3\font minus
		\fontdimen4\font\relax}
	\providecommand{\BIBforeignlanguage}[2]{{%
			\expandafter\ifx\csname l@#1\endcsname\relax
			\typeout{** WARNING: IEEEtran.bst: No hyphenation pattern has been}%
			\typeout{** loaded for the language `#1'. Using the pattern for}%
			\typeout{** the default language instead.}%
			\else
			\language=\csname l@#1\endcsname
			\fi
			#2}}
	\providecommand{\BIBdecl}{\relax}
	\BIBdecl
	
	\bibitem{4623708}
	V.~Chandrasekhar, J.~G. Andrews, and A.~Gatherer, ``Femtocell networks: a
	survey,'' \emph{IEEE Communications Magazine}, vol.~46, no.~9, pp. 59--67,
	September 2008.
	
	\bibitem{6768783}
	H.~Claussen, L.~T.~W. Ho, and L.~G. Samuel, ``An overview of the femtocell
	concept,'' \emph{Bell Labs Technical Journal}, vol.~13, no.~1, pp. 221--245,
	Spring 2008.
	
	\bibitem{6171992}
	J.~G. Andrews, H.~Claussen, M.~Dohler, S.~Rangan, and M.~C. Reed, ``Femtocells:
	Past, present, and future,'' \emph{IEEE Journal on Selected Areas in
		Communications}, vol.~30, no.~3, pp. 497--508, April 2012.
	
	\bibitem{anpalagan_bennis_vannithamby_2015}
	M.~B. A.~Anpalagan and R.~Vannithamby, \emph{Design and Deployment of Small
		Cell Networks}.\hskip 1em plus 0.5em minus 0.4em\relax Cambridge University
	Press, 2015.
	
	\bibitem{7839266}
	S.~Bassoy, H.~Farooq, M.~A. Imran, and A.~Imran, ``Coordinated multi-point
	clustering schemes: A survey,'' \emph{IEEE Communications Surveys Tutorials},
	vol.~19, no.~2, pp. 743--764, Secondquarter 2017.
	
	\bibitem{6530435}
	S.~S. Ali and N.~Saxena, ``A novel static clustering approach for {CoMP},'' in
	\emph{2012 7th International Conference on Computing and Convergence
		Technology (ICCCT)}, Dec 2012, pp. 757--762.
	
	\bibitem{6707857}
	H.~Shimodaira, G.~K. Tran, K.~Araki, K.~Sakaguchi, S.~Konishi, and S.~Nanba,
	``Diamond cellular network — optimal combination of small power
	basestations and {CoMP} cellular networks-,'' in \emph{2013 IEEE 24th
		International Symposium on Personal, Indoor and Mobile Radio Communications
		(PIMRC Workshops)}, Sept 2013, pp. 163--167.
	
	\bibitem{6181826}
	A.~Barbieri, P.~Gaal, S.~Geirhofer, T.~Ji, D.~Malladi, Y.~Wei, and F.~Xue,
	``Coordinated downlink multi-point communications in heterogeneous cellular
	networks,'' in \emph{2012 Information Theory and Applications Workshop}, Feb
	2012, pp. 7--16.
	
	\bibitem{6555174}
	A.~M. Hamza and J.~W. Mark, ``A clustering scheme based on timing requirements
	in coordinated base-stations cooperative communications,'' in \emph{2013 IEEE
		Wireless Communications and Networking Conference (WCNC)}, April 2013, pp.
	3764--3769.
	
	\bibitem{5594575}
	F.~Huang, Y.~Wang, J.~Geng, M.~Wu, and D.~Yang, ``Clustering approach in
	coordinated multi-point transmission/reception system,'' in \emph{2010 IEEE
		72nd Vehicular Technology Conference - Fall}, Sept 2010, pp. 1--5.
	
	\bibitem{5502468}
	S.~A. Ramprashad, G.~Caire, and H.~C. Papadopoulos, ``A joint scheduling and
	cell clustering scheme for mu-mimo downlink with limited coordination,'' in
	\emph{2010 IEEE International Conference on Communications}, May 2010, pp.
	1--6.
	
	\bibitem{6655533}
	V.~Pichapati and P.~Gupta, ``Practical considerations in cluster design for
	co-ordinated multipoint {(CoMP)} systems,'' in \emph{2013 IEEE International
		Conference on Communications (ICC)}, June 2013, pp. 5860--5865.
	
	\bibitem{4533793}
	A.~Papadogiannis, D.~Gesbert, and E.~Hardouin, ``A dynamic clustering approach
	in wireless networks with multi-cell cooperative processing,'' in \emph{2008
		IEEE International Conference on Communications}, May 2008, pp. 4033--4037.
	
	\bibitem{5285181}
	W.~Saad, Z.~Han, M.~Debbah, and A.~Hjorungnes, ``A distributed coalition
	formation framework for fair user cooperation in wireless networks,''
	\emph{IEEE Transactions on Wireless Communications}, vol.~8, no.~9, pp.
	4580--4593, September 2009.
	
	\bibitem{6786390}
	V.~Garcia, Y.~Zhou, and J.~Shi, ``Coordinated multipoint transmission in dense
	cellular networks with user-centric adaptive clustering,'' \emph{IEEE
		Transactions on Wireless Communications}, vol.~13, no.~8, pp. 4297--4308, Aug
	2014.
	
	\bibitem{8260866}
	Z.~Zhang, N.~Wang, J.~Zhang, and X.~Mu, ``Dynamic user-centric clustering for
	uplink cooperation in multi-cell wireless networks,'' \emph{IEEE Access},
	vol.~6, pp. 8526--8538, 2018.
	
	\bibitem{7008373}
	R.~Wei, Y.~Wang, and Y.~Zhang, ``A two-stage cluster-based resource management
	scheme in ultra-dense networks,'' in \emph{2014 IEEE/CIC International
		Conference on Communications in China (ICCC)}, Oct 2014, pp. 738--742.
	
	\bibitem{Tang2015}
	S.~Tang, C.~Sun, J.~Wang, Y.~Zhang, and F.~Wen, ``Interference management based
	on cell clustering in ultra-highly dense small cell networks,'' in \emph{2015
		International Conference on Information and Communications Technologies (ICT
		2015)}, April 2015, pp. 1--6.
	
	\bibitem{7498053}
	S.~Chen, C.~Xing, Z.~Fei, H.~Wang, and Z.~Pan, ``Dynamic clustering algorithm
	design for ultra dense small cell networks in {5G},'' in \emph{2015 10th
		International Conference on Communications and Networking in China
		(ChinaCom)}, Aug 2015, pp. 836--840.
	
	\bibitem{7794900}
	Z.~Xiao, J.~Yu, T.~Li, Z.~Xiang, D.~Wang, and W.~Chen, ``Resource allocation
	via hierarchical clustering in dense small cell networks: A correlated
	equilibrium approach,'' in \emph{2016 IEEE 27th Annual International
		Symposium on Personal, Indoor, and Mobile Radio Communications (PIMRC)}, Sept
	2016, pp. 1--5.
	
	\bibitem{8284755}
	W.~Yao, J.~Li, B.~Tan, and S.~Hao, ``Interference management scheme of ultra
	dense network based on clustering,'' in \emph{2017 IEEE 2nd Information
		Technology, Networking, Electronic and Automation Control Conference
		(ITNEC)}, Dec 2017, pp. 374--377.
	
	\bibitem{5963458}
	P.~Marsch and G.~Fettweis, ``Static clustering for cooperative multi-point
	{(CoMP)} in mobile communications,'' in \emph{2011 IEEE International
		Conference on Communications (ICC)}, June 2011, pp. 1--6.
	
	\bibitem{1237143}
	S.~Vishwanath, N.~Jindal, and A.~Goldsmith, ``Duality, achievable rates, and
	sum-rate capacity of gaussian {MIMO} broadcast channels,'' \emph{IEEE
		Transactions on Information Theory}, vol.~49, no.~10, pp. 2658--2668, Oct
	2003.
	
	\bibitem{5165179}
	J.~Papandriopoulos and J.~S. Evans, ``{SCALE}: A low-complexity distributed
	protocol for spectrum balancing in multiuser {DSL} networks,'' \emph{IEEE
		Transactions on Information Theory}, vol.~55, no.~8, pp. 3711--3724, Aug
	2009.
	
	\bibitem{414651}
	R.~D. Yates, ``A framework for uplink power control in cellular radio
	systems,'' \emph{IEEE Journal on Selected Areas in Communications}, vol.~13,
	no.~7, pp. 1341--1347, Sept 1995.
	
	\bibitem{BienTibshirani2011}
	J.~Bien and R.~Tibshirani, ``Hierarchical clustering with prototypes via
	minimax linkage,'' \emph{Journal of the American Statistical Association},
	vol. 106, no. 495, pp. 1075--1084, 2011.
	
\end{thebibliography}


%\printindex

\end{document}